\documentstyle[prl,aps,preprint,epsf,tighten]{revtex}
\begin{document}

\draft

\title{ Quasiparticle Spectrum Near the Gap Node Directions
in the Mixed State of D-wave Superconductors}

\author{ A. S. Mel'nikov}
\address{Institute for Physics of Microstructures,
        Russian Academy of Sciences\\
         603600, Nizhny Novgorod, GSP-105, Russia}

\date{\today}
\maketitle
\begin{abstract}
We report on a calculation of the quantized energy spectrum and
quasiparticle eigenfunctions for low lying excitations in the mixed state
of clean d-wave superconductors.  Our study is based on an approximate
analytical solution of the Bogolubov-de Gennes equations for both
rectangular and triangular flux lattices with one of the primitive
translations chosen parallel to the gap node direction. For excitations
with momenta close to a certain gap node we have obtained a set of
eigenfunctions which appear to be extended along the chosen gap node
direction and localized along the perpendicular one on a scale determined
by the intervortex distance. The periodic
superfluid velocity field induces a band structure in the spectrum, which
depends essentially on the vortex lattice geometry.
\end{abstract}
\narrowtext
Recently a great deal of attention has been devoted to the
nature of the quasiparticle states in isolated vortices and
vortex lattices in superconductors with anisotropic pairing.
This problem is of considerable importance since
low-energy quasiparticle excitations
influence strongly on various static and dynamic
properties of the mixed state at low temperatures.
These investigations were stimulated by
a large number of experiments which provide good
evidence for the d-wave symmetry of the order parameter
in high-temperature superconductors
(see \cite{dwave} and references therein).
For conventional s-wave superconductors the low-lying
quasiparticle states are bound to the vortex cores as it
was first predicted by Caroli, de Gennes, and Matricon \cite{degen}.
For an isolated vortex line
the eigenvalues for these localized states may be written
as follows: $E_{\mu}\sim \mu \Delta/(k_F\xi)$, where
$\Delta$ is the gap value far from the vortex axis,
$\xi$ is the coherence length at $T=0$, $k_F$ is the Fermi momentum,
and the angular momentum quantum number $\mu$ is
half an odd integer.

In contrast to conventional superconductors, the density of states (DOS)
at low energies in d-wave systems is dominated by contributions
which come from the regions far from the vortex cores and depends
essentially on the intervortex distance $R_v$.
This unusual fact is a direct consequence of
the vanishing pair potential in directions of
the gap nodes. As a result,
one obtains the specific magnetic field dependence of
DOS at the Fermi level (see Refs.~\cite{volovik,barash,hirsch}), which
may be experimentally identified, for instance, in specific heat
\cite{moler,revaz}
or scanning tunneling microscope (STM) \cite{tunnel} measurements.
It should be also emphasized that there is an important difference
between the quantization of the energy spectrum in the mixed state of s-
and d-wave superconductors: for s-wave systems the energy quantization
and corresponding localized states exist even in a single isolated vortex
line (and they are weekly influenced by the presence of neibouring
vortices at least for $R_v\gg\xi$) while for the d-wave case the low
lying energy spectrum may be quantized only due to the finite
intervortex distance (see Refs.\cite{kv1,kv2,kopnin,makh,lee}).
The latter conclusion is proved by the numerical
solution of the Bogolubov-de Gennes (BdG) equations \cite{franz} which
shows that there are no truly localized states for a single isolated
vortex in a pure d-wave superconductor.

The most simple theoretical description of the excitation
spectrum is based on the semiclassical approach which
takes account of the Doppler shift
of the quasiparticle energy by the
local superfluid velocity ${\bf V}_s$:

\begin{equation}
\label{semiclass}
\varepsilon ({\bf k},{\bf r})=
\pm\sqrt{\hbar^2 V_F^2(k-k_F)^2+\Delta^2({\bf k})}
+\hbar{\bf k}{\bf V}_s({\bf r})
\end{equation}

where $V_F$ is the Fermi velocity.
Hereafter we assume the Fermi surface to be
two-dimensional, which is appropriate to high-$T_c$ superconductors,
and take the gap function in the form
$\Delta_d=\Delta_0 k_xk_y/k_F^2$ (the $x$ axis is taken along the [110]
crystal direction and thus makes an angle $\pi/4$ with the $a$ axis
of the $CuO_2$ planes).
We restrict ourselves to the study of a vortex lattice under
a magnetic field applied parallel to the $\it c$ axis and,
as a consequence, the vector ${\bf V}_s$ is a periodic
function of $x,y$ with the periodicity of the vortex lattice.
One can separate two length scales for quasiparticle
wavefunctions: an atomic lengh scale $k_F^{-1}$ and a
characteristic wavelength of a slowly varying envelope ${\it l}$.
The semiclassical procedure is in general correct when the order parameter
and superfluid velocity are modulated on a scale
$\Lambda\gg{\it l}$ .
For s-wave superconductors we have
${\it l}\sim\xi$ (the minimum spatial extension of wave packets made with
excitations (\ref{semiclass}))
and, as a result, the semiclassical approach
fails only for the vortex core region. For d-wave systems the
${\it l}$ value appears to be angular dependent.
In the homogeneous case the spectrum for the low-lying excitations which
are close to one of the gap nodes (which corresponds, e.g., to the
point ${\bf k}=(k_F,0)$) has the form:

\begin{equation}
\label{enode}
\varepsilon ({\bf k})\simeq\pm\Delta_0
\sqrt{\xi^2 (k-k_f)^2+k_y^2/k_F^2},
\end{equation}

where $\xi=\hbar V_F/\Delta_0$.
The typical momenta in the $x$ direction are
$q_x\sim\varepsilon/(\Delta_0\xi)$ and the corresponding wavelength is of
the order ${\it l}_x\sim \Delta_0\xi/\varepsilon$.
Comparing the ${\it l}_x$ value
with the intervortex distance $R_v$ (the characteristic length
of the superfluid velocity variation), one can expect
that the semiclassical approach based on the
expression (\ref{semiclass}) fails
for low energies $\varepsilon\stackrel{_<}{_\sim}\Delta_0\xi/R_v$.
In this case we may assume that the effective potential
$\hbar{\bf k}{\bf V}_s({\bf r})$ in Eq.~(\ref{semiclass})
should be averaged over the distances $\sim {\it l}_x$ in the $x$
direction (the validity of these qualitative arguments will be proved
below).
To analyse the problem beyond the
semiclassical approach one must use the more
powerful methods based on either the BdG equations
or Green's-function techniques (which are equivalent for clean
superconductors).
The quasiclassical limit
($k_F\xi\gg 1$) of these theories is known to be represented by
the Andreev's and Eilenberger equations, respectively.
Within these models one should solve the one-dimensional quantum
mechanical problem for the particle motion along the quasiclassical
trajectory which is characterized by the impact parameter $b=\mu/k_F$
and the angle $\theta$ in the $x-y$ plane
\cite{volovik,kv1,kv2,kopnin,makh,maki1,ichioka}. Using the
the Bohr-Sommerfeld quantization rule for the angular momentum
$\mu(\theta,\varepsilon)$ one can in principle determine the true
quantum levels (see Refs.~\cite{kv1,kv2,kopnin,makh}).
For the d-wave case the main difficulty is connected
with the description of quasiclassical trajectories which make
an angle $\theta< \xi/R_v$ with the gap node directions.
For this angular domain the extension of the
wave function exceeds the $R_v$ value and quasiparticle
states are sensitive to
the superfluid velocity fields of all vortices
even if the impact parameter is less than $R_v$.
The correct description of these trajectories with small $\theta$ values
is of considerable importance since it is this angular interval
which determines the true quantum levels according to the Bohr-Sommerfeld
quantization rule.

It is the purpose of this Letter to report on a calculation of the
quantized energy spectrum and quasiparticle eigenfunctions for various
vortex lattice structures, based on an approximate analytical solution of
the BdG equations for low lying excitations with
$\varepsilon\stackrel{_<}{_\sim}\Delta_0\xi/R_v$.
Our approach provides a possibility to take proper account of the
influence of the superfluid velocity fields of all vortices
in the vortex lattice on the quasiparticle motion along the trajectories
almost parallel to gap node directions.
We consider the case of pure d-wave superconductivity and
neglect the effects connected with the possible
formation of states of mixed symmetry
(with coexisting s- and d-wave or $d_{x^2-y^2}$ and $d_{xy}$
order parameter components).
The BdG equations can be written as:

\begin{eqnarray}
\label{bdg1}
\hat h_0 u({\bf r})+\int \Delta({\bf r},{\bf r}^\prime)
v({\bf r}^\prime)d{\bf r}^\prime=\varepsilon u({\bf r})\\
\label{bdg11}
-\hat h_0^* v({\bf r})+\int \Delta^*({\bf r},{\bf r}^\prime)
u({\bf r}^\prime)d{\bf r}^\prime=\varepsilon v({\bf r})
\end{eqnarray}

where $u,v$ are the particlelike and holelike parts of the
quasiparticle wave function.
The one-particle Hamiltonian $\hat h_0$
in the most simple isotropic case takes the form
$$
\hat h_0=-\frac{\hbar^2}{2m}
\left(\nabla +i\frac{\pi}{\phi_0}{\bf A}\right)^2-E_F,
$$
where $\phi_0$ is the flux quantum and $E_F$ is the Fermi energy.
The system (\ref{bdg1},\ref{bdg11})
 has been previously solved numerically
for specific lattice models \cite{soin,wang} and in the continuum limit
\cite{franz,maki118}. To obtain the analytical solution we
follow the well known procedure
(see Refs.~\cite{lee,bruder}) and
simplify the nonlocal off-diagonal terms using the condition
$k_F\xi\gg 1$ (or, equivalently, $\Delta_0\ll E_F$).
In this case one can search for the solution in the form
$u=Uexp(i{\bf k}_F{\bf r})$, $v=Vexp(i{\bf k}_F{\bf r})$,
i.e. divide out the fast oscillations on a scale
$k_F^{-1}$. Then we rewrite $\Delta({\bf r},{\bf r}^\prime)$
in terms of the center of mass
${\bf R}=({\bf r}+{\bf r}^\prime)/2$ and relative coordinates
${\bf \rho}={\bf r}-{\bf r}^\prime$ and introduce the gap function
as a Fourier transform with respect to ${\bf \rho}$:
$\Delta({\bf k},{\bf R})=\Delta_d({\bf k})\Psi({\bf R})$.
The function $\Psi({\bf R})=fexp(i\chi)$ is the
d-wave order parameter used in Ginzburg-Landau theory.
Let us take the two-term expansion for the gap
operator:

\begin{equation}
\int \Delta({\bf r},{\bf r}^\prime)
v({\bf r}^\prime)d{\bf r}^\prime\simeq
e^{i{\bf k}_F{\bf r}}
\left(\left.
\Psi\Delta_d({\bf k}_F)V-
\frac{i}{2}\frac{\partial\Delta_d}{\partial {\bf k}}\right|_{{\bf k}_F}
\{\nabla,\Psi\}V\right) ,
\end{equation}

where we use the notation $\{\hat A,\hat B\}$ for the anti-commutator
of two operators $\hat A$ and $\hat B$.
In order to obtain the Andreev's equations we should keep only
the first term in this expression.
Obviously such an approximation is not correct
and the second term can not be omitted
when ${\bf k}_F$ is close to the
gap nodes and the first term vanishes.
Taking, e.g., ${\bf k}_{F1}=(k_F,0)$ and introducing
a new two-component wave function
$\hat g=(Uexp(-i\chi),V)$ (to eliminate the order parameter phase
in $\Psi$) one obtains the equations linearized in gradient terms:

\begin{eqnarray}
-\xi\hat\sigma_z \left(i\frac{\partial}{\partial x}+
\frac{\pi A_x}{\phi_0}  \right)\hat g
-\frac{ \hat\sigma_x}{2k_F}
\left\{i\frac{\partial}{\partial y}+\frac{\pi A_y}{\phi_0},f\right\}\hat g
\nonumber\\
\label{bdg2}
+\frac{m\xi}{\hbar}V_{sx}(1+\hat\sigma_z) \hat g+
\frac{m f}{\hbar k_F}  V_{sy}\hat\sigma_x \hat g
=\frac{\varepsilon}{\Delta_0}\hat g,
\end{eqnarray}

where $\hat\sigma_x,\hat\sigma_y, \hat\sigma_z$ are the Pauli matrices.
We follow here the treatment in Ref.~\cite{lee}
and neglect the curvature of the Fermi surface.
For an isotropic Fermi surface
such an approximation is valid only
for $\varepsilon\ll \Delta_0/(k_F\xi)$
(see Refs.~\cite{kvcom,leerep}). However, as mentioned in
\cite{lee,leerep}, the range of validity of Eq.~(\ref{bdg2}) may be even
larger if the Fermi surface is somewhat flattened at the
nodes.
At intermediate magnetic fields $H_{c1}\ll H\ll H_{c2}$
(when the intervortex distance is much smaller than the
London penetration depth)
we can assume ${\bf H}=-H{\bf z}_0$ to be homogeneous and take
the gauge ${\bf A}=Hy{\bf x}_0$.
Here ${\bf x}_0$, ${\bf y}_0$, ${\bf z}_0$
are the unit vectors of the coordinate system.
For a vortex lattice with primitive translations ${\bf a}_1,{\bf a}_2$
the superfluid velocity may be written in the form:

\begin{equation}
\label{velocity}
{\bf V}_s=\frac{i\pi\hbar H}{m\phi_0}
\sum_{{\bf b}\neq 0}
\frac{[{\bf b},{\bf z}_0]}{|{\bf b}|^2}
e^{i{\bf b}{\bf r}},
\end{equation}

where ${\bf b}=n{\bf b}_1+m{\bf b}_2$
and ${\bf b}_1=2\pi H[{\bf a}_2,{\bf z}_0]/\phi_0$,
${\bf b}_2=2\pi H[{\bf z}_0,{\bf a}_1]/\phi_0$
are the primitive translations in the reciprocal lattice,
$n$ and $m$ run over all possible integers.
As we see below the solution of Eq.~(\ref{bdg2}) depends
strongly on the flux lattice structure and its orientation
relative to the crystal axes.
Contrary to the conventional isotropic superconductors
(where a hexagonal flux lattice appears to be energetically
favourable), for d-wave compounds previous theoretical works
predicted a rich phase diagram, containing triangular, centered
rectangular and square lattices with various orientations relative to the
ionic lattice, as a function of magnetic field and temperature
\cite{amin1,amin2,makisq,maki310,maki067,ich147}.
In principle one should treat
the problem self-consistently, i.e. calculate at first
the energy spectra for various vortex configurations and then
find the lattice structure corresponding to the free energy minimum.

In this paper we do not solve this self-consistent problem and
restrict ourselves to the study of energy spectra for
several particular lattice structures.
Let us choose one of the primitive translations (e.g. ${\bf a}_1$)
to be parallel to the gap node direction ${\bf x}_0$
and consider two types of lattices:
(I) ${\bf a}_1=a{\bf x}_0$, ${\bf a}_2=\sigma a{\bf y}_0$,
$H\sigma a^2=\phi_0$ (rectangular lattice);
(II) ${\bf a}_1=a{\bf x}_0$, ${\bf a}_2=a ({\bf x}_0/2-\sigma {\bf y}_0)$,
$H\sigma a^2=\phi_0$
(vortices in the unit cell form a shape of an isosceles triangle
with the base along the x axis).
Note that the centered rectangular lattices of the type II
(with the gradually changing parameter $\sigma$
as a function of $H$ and $T$ values)
were found to be energetically favourable
for a certain region of the $H-T$ phase diagram
within a generalized London model taking account of nonlocal and
nonlinear corrections to the free energy \cite{amin1,amin2}.
We also include in our consideration
the square lattice tilted by $\pi/4$ from the $a$ axis
(type I, $\sigma=1$) which is
most stable at least for rather high fields
and temperatures not very close to $T_c$ according to
Refs.~\cite{makisq,maki310,maki067,ich147}.
Such a lattice structure (though elongated in the $a$ direction) is
close to the one observed experimentally in twinned YBaCuO monocrystals
by small angle neutron scattering \cite{neutron} and scanning tunneling
microscopy \cite{tunnel}.

If we search for the solution of Eq.~(\ref{bdg2}) as
a linear combination of harmonics $e^{iq_x x} \hat G(q_x,y)$,
the periodic functions $f$ and ${\bf V}_s$ will be responsible
for the interaction of harmonics with $q_x$ and $q_x+nb_{1x}$.
As we see below, for $\varepsilon<0.5\pi\Delta_0\xi/a$ the wavefunctions
$\hat G(q_x,y)$ and $\hat G(q_x+nb_{1x},y)$ do not overlap in the $y$
direction and, consequently, their interaction is negligible.  In this
case the $q_x$ momentum component is a good quantum number and  one can
replace the exact periodic potential (\ref{velocity}) in Eq.~(\ref{bdg2})
by the effective potential averaged in the $x$ direction.  The equation
for $\hat G(q_x,y)$ reads:

\begin{equation}
\label{bdg3}
\xi\hat\sigma_z \left(q_x-
\frac{\pi Hy}{\phi_0}  \right)\hat G
-\frac{ i}{k_F} \hat\sigma_x
\frac{\partial\hat G}{\partial y}
+\frac{\pi\xi}{2}\sqrt{\frac{\sigma H}{\phi_0}}
\Phi\left(y\sqrt{\frac{H}{\sigma\phi_0}}\right)
(1+\hat\sigma_z)\hat G
=\frac{\varepsilon}{\Delta_0}\hat G,
\end{equation}

where $\Phi(z)=2z-(2m+1)$ for $m<z<m+1$, $m\in Z$.
We omitted here the small corrections of the order $\xi/a$,
which are connected with the suppression of the order parameter $f$
in vortex cores.
It is convinient to introduce
$\hat F=(\hat\sigma_x+\hat\sigma_z)\hat G/2$
and dimensionless variables $y\sqrt{H/(\sigma\phi_0)}=z$,
$q_x=\pi Q\sqrt{H\sigma/\phi_0}$,
$\varepsilon=E\Delta_0\pi \xi\sqrt{H\sigma/\phi_0}$.
At the $m$-th interval ($m<z<m+1$) one obtains:

\begin{equation}
\label{bdg4}
-i\lambda\hat\sigma_z\frac{\partial \hat F}{\partial z}+
\left(\frac{1}{2}\Phi(z)-E\right)\hat F+
q_m\hat\sigma_x \hat F=0,
\end{equation}

where $q_m=Q-m-1/2$ and $\lambda=(\pi\sigma k_F\xi)^{-1}$ is a
dimensionless wavelength.
The procedure analogous to the one used above for the gap node
at the point ${\bf k}_{F1}=(k_F,0)$ can be carried out for
the perpendicular gap node direction ${\bf k}_{F2}=(0,k_F)$.
Taking the gauge ${\bf A}=-Hx{\bf y}_0$ and introducing
$\hat F=(\hat\sigma_x+\hat\sigma_z)\hat G^*/2$,
$-x\sqrt{H\sigma/\phi_0}=z$,
$q_y=\pi Q\sqrt{H/(\sigma\phi_0)}$,
$\varepsilon=E\Delta_0\pi \xi\sqrt{H/(\sigma\phi_0)}$,
$\lambda=\sigma(\pi k_F\xi)^{-1}$
($-2x\sqrt{H\sigma/\phi_0}=z$,
$2q_y=\pi Q\sqrt{H/(\sigma\phi_0)}$,
$2\varepsilon=E\Delta_0\pi \xi\sqrt{H/(\sigma\phi_0)}$,
$\lambda=4\sigma(\pi k_F\xi)^{-1}$ )
for the type I (II) lattices one obtains Eq.~(\ref{bdg4}).
The symmetry of BdG equations allows us to conclude
that the solutions $U_{3,4},V_{3,4},\varepsilon_{3,4}$ for two other gap
nodes at the points ${\bf k}_{F3}=(-k_F,0)$ and ${\bf k}_{F4}=(0,-k_F)$
can be found by simple transformation:
$U_{3,4}=V^*_{1,2}$, $V_{3,4}=-U^*_{1,2}$,
$\varepsilon_{3,4}=-\varepsilon_{1,2}$.

It may be useful to note that inside the $m$-th interval
the equations (\ref{bdg4}) are equivalent to
the ones describing the interband tunneling
\cite{kane} or the one-dimensional
(1D) motion of a Dirac particle
in a uniform electric field and can be solved exactly in terms of
parabolic cylinder functions. To obtain the energy spectrum
one should match these solutions at points $z=m$.
This procedure may be essentially simplified
using the usual 1D quasiclassical approach which is valid
for $\lambda\ll 1$.
Substituting $\hat F\propto exp(iS(z))$ in Eq.~(\ref{bdg4})
we obtain the classically allowed (CA) regions (where the function $S$ is
real):  $|\Phi(z)/2-E|>|q_m|$.  These regions shift
in the $z$ direction with a change in the $Q$ value and the
quasiparticle motion at the $m$-th interval is classically allowed only
for $|q_m|<1/2+|E|$.
From the latter condition one can see that for rather low energies
($|E|<1/2$) the CA regions corresponding to the values $Q$ and $Q+2n$ do
not overlap. In this case the interaction of harmonics with $q_x$ and
$q_x+nb_{1x}$ will be exponentially small
due to the exponential decay of the wavefunction
in classically forbidden regions.
Thus, our procedure based on the averaging of the superfluid velocity
in the $x$ direction is correct only for low energies $|E|<1/2$,
when the solutions are essentially localized
in the $y$ direction on a scale determined by the intervortex
distance.

Let us now consider the domain $|q_m|<1$ and
continue with the calculation of
quasiclassical wavefunctions and energy levels corresponding
to the quasiparticle motion at the $m$-th interval which
contains two CA regions: (i) $-1/2<z-m-1/2<E-|q_m|$ (region A)
and (ii) $E+|q_m|<z-m-1/2<1/2$ (region B).
Note that the region A (B) exists
if the turning point $z_{1m}=m+1/2+E-|q_m|$
($z_{2m}=m+1/2+E+|q_m|$)
belongs to the $m$-th interval.
The wavefunctions and quantization rules for regions A and B
can be written as follows:

\begin{eqnarray}
\hat F_{A,B}=\frac{C_{A,B}}{(t^2-q_m^2)^{1/4}}
\left(\frac{exp(iS_{A,B})}{|t+\sqrt{t^2-q_m^2}|^{1/2}}
\left(-q_m\atop t+\sqrt{t^2-q_m^2}\right)\right.
+ \nonumber\\
\label{sol}
\left.+\frac{exp(-iS_{A,B})}{|t-\sqrt{t^2-q_m^2}|^{1/2}}
\left(-q_m\atop t-\sqrt{t^2-q_m^2}\right)\right)\\
S_{A,B} =\frac{1}{\lambda}
\int\limits_{\mp |q_m|}^t\sqrt{s^2-q_m^2}ds
\pm \frac{\pi}{4}\nonumber\\
\label{qr1}
\int\limits_{|q_m|}^{1/2\pm E}\sqrt{t^2-q_m^2}dt=
\pi\lambda (N_{A,B}+\gamma_{A,B}(q_m))
\end{eqnarray}

where $t=z-m-1/2-E$, $N_{A,B}$ is an integer.
The $\gamma_{A,B}$ values are of the order unity and determined
by the matching of the expression (\ref{sol}) with the exponentially
decaying solutions solutions in classically forbidden regions at the
$(m-1)$-th and $(m+1)$-th intervals. This matching procedure results in
the following boundary conditions
for the wavefunction $\hat F=(F_1,F_2)$:

\begin{eqnarray}
\label{bc1}
\left.\frac{F_2}{F_1}\right|_{z=m}=exp(i\alpha_A);
\qquad cos\alpha_A=\frac{E-1/2}{Q-m+1/2}\\
\label{bc2}
\left.\frac{F_2}{F_1}\right|_{z=m+1}=exp(i\alpha_B);
\qquad cos\alpha_B=\frac{E+1/2}{Q-m-3/2}
\end{eqnarray}

Evaluating the integral in the l.h.s. of Eq.~(\ref{qr1})
one obtains:

\begin{equation}
\label{qr2}
\left(\frac{1}{2}\pm E_{A,B}\right)
\sqrt{\left(\frac{1}{2}\pm E_{A,B}\right)^2-q_m^2}-
q_m^2 cosh^{-1}\left(\frac{\frac{1}{2}\pm E_{A,B}}{|q_m|}\right)=
2\pi\lambda (N_{A,B}+\gamma_{A,B})
\end{equation}

\begin{figure}[h]
\leavevmode
\epsfysize=21cm
\epsfbox{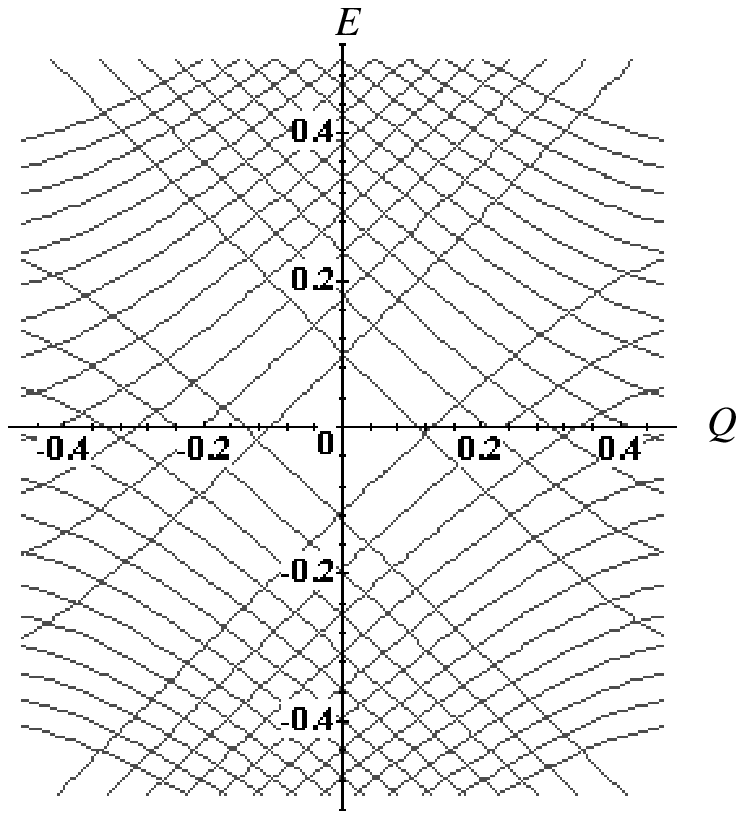}
\caption{
The energy branches in the first
Brillouin zone for $\lambda=0.01$.
}
\label{fig1}
\end{figure}

The important point is that a set of eigenvalues corresponding
to a certain momentum $Q$ coincides with
that for the momentum $Q+1$. Such a periodicity of
the energy spectrum is a consequence of
the periodicity of the potential $\Phi(z)$
and can be proved exactly from Eq.~(\ref{bdg4}).
Thus, to analyse the spectrum one can consider only the $Q$ values in the
1D Brillouin zone $-1/2<Q<1/2$.
In Fig.~\ref{fig1} we display the energy branches (\ref{qr2}) in the first
Brillouin zone for the particular case $\lambda=0.01$.
Each energy branch $E_A(m,N_A,Q)$ ($E_B(m,N_B,Q)$) considered
as a function of the momentum $Q$ has a single minimum (maximum)
at the point close to the Brillouin zone boundary $Q=m+1/2$.
Near the first Brillouin zone boundary $|Q-1/2|\ll 1/2\pm E$ we obtain:

\begin{equation}
\label{qr3}
\left.E_{A,B}\right|_{m=0}
\simeq \mp \frac{1}{2}\pm \sqrt{2\pi\lambda
(N_{A,B}+\gamma_{A,B})}\pm \frac{(Q-1/2)^2}{2\sqrt{2\pi\lambda
(N_{A,B}+\gamma_{A,B})}} ln \frac{\sqrt{2\pi\lambda
(N_{A,B}+\gamma_{A,B})}}{|Q-1/2|}
\end{equation}

Near the points of intersection of the energy branches (\ref{qr2})
in the $E-Q$ plane one should take account of the splitting
of energy levels resulting from the quasiparticle tunneling
through classically forbidden regions.
As a result we obtain a set of narrow bands separated by energy gaps.
The effect of tunneling (and, therefore, energy splitting) is
exponentially small if the characteristic length of the wavefunction decay
in a classically forbidden region is much smaller than
the dimension of this region.
In the opposite case the effect of tunneling is essential.
In particular, we can not
neglect the tunneling
between the regions A and B for $Q$ close to Brillouin zone boundaries,
when the distance between the turning points $z_{1m}$ and $z_{2m}$
becomes comparable to the characteristic decay length
$\sqrt{\lambda}$ in the region $z_{1m}<z<z_{2m}$.
To consider this limit let us take the exact solution at the
$m$-th interval:

\begin{equation}
\label{exact}
\hat F=\left(AD_{-i\mu-1}(\tau)+BD_{-i\mu-1}(-\tau)
\atop
sign(Q-m-1/2)\sqrt{i/\mu} (-AD_{-i\mu}(\tau)+
BD_{-i\mu}(-\tau)) \right)
\end{equation}

Here $\tau=\frac{2}{i\lambda}(z-m-1/2-E)$,
$\mu=q_m^2/(2\lambda)$,
$D_{-i\mu-1}(\tau)$ and $D_{-i\mu}(\tau)$ are the parabolic
cylinder functions \cite{witt}.
Using Eqs.~(\ref{bc1},\ref{bc2},\ref{exact}) one obtains the system of two
equations for $A$ and $B$. The solvability condition for this system
results in the quantization rule.  Far from the turning points the
parabolic cylinder functions can be replaced by the corresponding
asymptotic expressions.  Thus, if the turning points $z_{1m}$ and
$z_{2m}$ are not too close to the $m$-interval boundaries ($z_{1m}-m\gg
|q_m|$, $m+1-z_{2m}\gg |q_m|$) and the distance $z_{2m}-z_{1m}$
is rather small ($|q_m|\ll\sqrt{\lambda}$) then
the quantization rule takes the following form:

\begin{equation}
\label{qr4}
E\simeq E_{0N}+ \sqrt{\pi\lambda}q_m (-1)^N
sin\left(\frac{\pi}{4}+\frac{1}{4\lambda}+
\frac{E_{0N}^2}{4\lambda}-
\frac{q_m^2}{2\lambda}ln\frac{(1/4-E_{0N}^2)}{\lambda}-
\frac{\alpha_A+\alpha_B}{2}\right)
\end{equation}
$$
E_{0N}=\pi N\lambda+\lambda(\alpha_A-\alpha_B)/2
$$
where $N\in Z$.
The spectrum appears to be equidistant only at the
Brillouin zone boundaries ($Q=\pm 1/2$) and the distance between the
energy levels depends strongly on the vortex lattice geometry.

Using the solution of Eq.~(\ref{bdg4}) given above we obtain four
different sets of eigenfunctions and eigenvalues associated with
four gap nodes. The range of validity of our approach is restricted by
the conditions $\varepsilon<
0.5\Delta_0\pi\xi\sqrt{H/\phi_0}min[\sigma^{-1/2},\sigma^{1/2}]$
and $\varepsilon<
0.5\Delta_0\pi\xi\sqrt{H/\phi_0}min[\sigma^{-1/2}/2,\sigma^{1/2}]$
for type I and II lattices, respectively.
In the opposite limit the wavefunctions are not localized in
the $z$ direction and our approximate solution based on the averaging of
the superfluid velocity in the gap node direction is no more valid.
Thus, for quasiparticle states associated with each gap node
${\bf k_{F\alpha}}$ there exist two regimes with the crossover parameter
$\varepsilon R_{v\alpha}/(\Delta_0\xi)$ (where $R_{v\alpha}$ is the
intervortex distance in the gap node direction, $\alpha=x,y$): (i) for
the low energy regime the spectrum has a band structure
and wavefunctions are extended in the gap node direction and localized in
the perpendicular one; (ii) for the high energy regime
the spectrum is continuous and eigenfunctions are extended in both
directions.  The characteristic energy $\varepsilon$ of excitations
coming into play at finite temperatures is of the order $\sim T$ and,
consequently, for thermodynamic and transport quantities one can expect
different regimes with the crossover parameters $T R_{vx}/(\Delta_0\xi)$,
$T R_{vy}/(\Delta_0\xi)$.  Taking the case $R_{vx}\sim R_{vy}\sim
\sqrt{\phi_0/H}$ we obtain the crossover parameter introduced previously
in Refs.~\cite{kv1,kopnin,kvcom}.

\begin{figure}[h]
\leavevmode
\epsfysize=21cm
\epsfbox{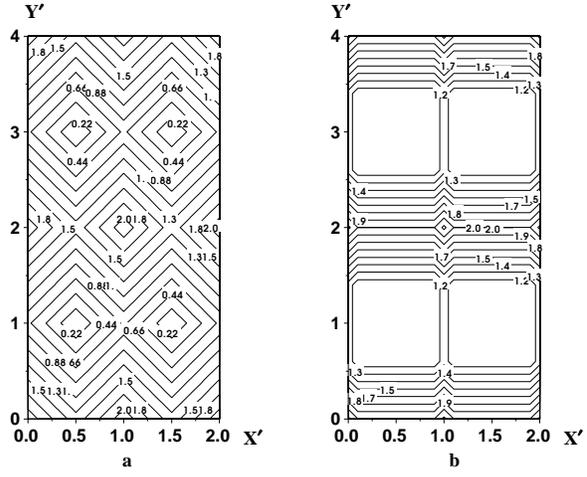}
\caption{Contour plot of
the local DOS
for the rectangular lattice with
$\sigma=2$
($x^\prime=x(2H/\phi_0)^{0.5}$, $y^\prime=y(2H/\phi_0)^{0.5}$):
(a)$\varepsilon=0$;
(b)$\varepsilon=0.2\pi\xi\Delta_0(2H/\phi_0)^{0.5}$.
}
\label{fig2}
\end{figure}

\begin{figure}[h]
\leavevmode
\epsfysize=21cm
\epsfbox{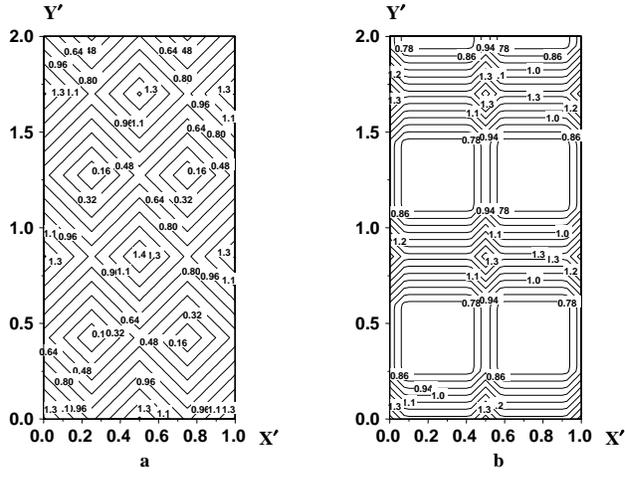}
\caption{Contour plot of
the local DOS
for the hexagonal lattice:
($x^\prime=x(\sigma H/\phi_0)^{0.5}$,
 $y^\prime=y(\sigma H/\phi_0)^{0.5}$):
(a)$\varepsilon=0$;
(b)$\varepsilon=0.2\pi\xi\Delta_0(\sigma H/\phi_0)^{0.5}$.}
\label{fig3}
\end{figure}

The quantization of the low energy spectrum should result in
the oscillatory behavior of DOS as a function of energy with the
characteristic energy scales of the order $\delta \varepsilon_1 \sim
\Delta_0\sqrt{\hbar \omega_c/(\sigma E_F)}$ and
$\delta \varepsilon_2 \sim \Delta_0\sqrt{\hbar \omega_c\sigma/E_F}$,
where $\omega_c=eH/(mc)$ is the cyclotron frequency.
Even if we neglect these oscillations
and assume $N$ to be a continuous variable
(more detailed study is left as a future problem)
the spatial and energy
dependence of the local DOS
 in the low energy regime
 differs qualitatively from the one obtained
from the semiclassical approach based on the expression (\ref{semiclass}).
Taking account of contributions from four gap nodes we obtain:

\begin{eqnarray}
N_I(\varepsilon,x,y)=\frac{N_F}{4}
\sqrt{\frac{\pi H}{2H_{c2}}}\left(
\frac{1}{\sqrt{\sigma}}
f\left(x\sqrt{\frac{H\sigma}{\phi_0}},
\frac{\varepsilon}{\pi\xi\Delta_0}\sqrt{\frac{\sigma\phi_0}{H}}\right)
+\sqrt{\sigma}
f\left(y\sqrt{\frac{H}{\sigma\phi_0}},
\frac{\varepsilon}{\pi\xi\Delta_0}\sqrt{\frac{\phi_0}{\sigma H}}\right)
\right)\\
N_{II}(\varepsilon,x,y)=\frac{N_F}{4}
\sqrt{\frac{\pi H}{2H_{c2}}}\left(
\frac{1}{2\sqrt{\sigma}}
f\left(2x\sqrt{\frac{H\sigma}{\phi_0}},
\frac{2\varepsilon}{\pi\xi\Delta_0}\sqrt{\frac{\sigma\phi_0}{H}}\right)
+\sqrt{\sigma}
f\left(y\sqrt{\frac{H}{\sigma\phi_0}},
\frac{\varepsilon}{\pi\xi\Delta_0}\sqrt{\frac{\phi_0}{\sigma H}}\right)
\right)
\end{eqnarray}
$$
f(z,\tilde\varepsilon)=|\Phi(z)/2-\tilde\varepsilon|+
|\Phi(z)/2+\tilde\varepsilon|
$$

where $N_I$ ($N_{II}$) is the local DOS for the lattices of the
type I (II), $N_F$  is the density of states at the Fermi level in a
normal state.
Contour plots of the functions
$4 N_I(x,y)N_F^{-1}\sqrt{2H_{c2}/(\pi H)}$
(for the rectangular lattice with $\sigma=2$)
and
$4 N_{II}(x,y)N_F^{-1}\sqrt{2H_{c2}/(\pi H)}$
(for the hexagonal lattice with $\sigma=\sqrt{3}/2$)
are shown in Fig.~\ref{fig2} and Fig.~\ref{fig3}, respectively.
The peaks in the local DOS occur at the points of
intersection of lines  parallel to the $x$ and $y$ axes and
passing through the vortex centers. The interesting fact
is that for type II lattices the peaks in the local DOS appear not only
at the vortex centers while for rectangular lattices (type I) the
coordinates of all peaks coincide with the vortex positions.  In the
vicinity of each vortex center ($\bar x,\bar y$) the local DOS exhibits a
fourfold symmetry which is in a good agreement with the previous studies
of a single isolated vortex in a d-wave superconductor
\cite{maki1,ichioka} and decreases linearly with an increase of
distances $|x-\bar x|$ and $|y-\bar y|$
from the vortex center
(contrary to the $((x-\bar x)^2+(y-\bar y)^2)^{-1/2}$ divergence
resulting from the semiclassical approach
based on Eq.~(\ref{semiclass})). The spatially averaged DOS
has the form

\begin{eqnarray}
<N_I> =\frac{N_F}{8} \sqrt{\frac{\pi H}{2H_{c2}}}
\left(\sqrt{\sigma}+\frac{1}{\sqrt{\sigma}}\right)
\left(1+\frac{4\varepsilon^2\phi_0}{\pi^2\xi^2\Delta_0^2 H} \right)
\\
<N_{II}> =\frac{N_F}{8} \sqrt{\frac{\pi H}{2H_{c2}}}
\left(\sqrt{\sigma}+\frac{1}{2\sqrt{\sigma}}\right)
\left(1+\frac{8\varepsilon^2\phi_0}{\pi^2\xi^2\Delta_0^2 H} \right)
\end{eqnarray}

If we assume the parameter $\sigma$ to be field independent then
the magnetic field dependence of the residual
DOS (at $\varepsilon=0$) follows the square root behavior which was first
predicted in Ref.~\cite{volovik}. The deviations from this behaviour may
appear if the vortex lattice structure and, hence, the $\sigma$ value
change with a magnetic field increasing
(see Refs.~\cite{amin1,amin2,maki310,maki067,ich147}) and, probably,
they are the cause for the discrepancies in specific heat measurements
\cite{moler,revaz}.
Note that these discrepancies may be
also explained taking account of the disorder effects
\cite{barash,hirsch}.

In summary, we have described the distinctive features of the low lying
quasiparticle states in the vortex lattices in d-wave systems.
The periodic
superfluid velocity field is shown to induce the band structure
in the low energy part of the spectrum.
This band structure and corresponding eigenfunctions have been
analysed for both rectangular and centered rectangular flux lattices
tilted by $\pi/4$ from the $a$ axis.
The resulting peculiarities of the
local DOS are shown to be strongly influenced by the vortex lattice
geometry.
The unusual behaviour of DOS discussed above can
be probed by specific heat and STM measurements and could provide
additional arguments in favour of the d-wave pairing in high-$T_c$
superconductors.
The present study provides a starting point
for the analysis of static and dynamic properties of
the mixed state in various d-wave systems, including, probably,
high-$T_c$ copper oxides.
For this purpose the approach developed above should be generalized to
describe the quasiparticle states for arbitrary lattice orientations with
respect to crystal axes.

It is my pleasure to thank Dr.I.D.Tokman
for very useful discussions of the issues considered in this article.
This work was supported, in part, by the Russian Foundation for
Fundumental Research (Grant No. 97-02-17437).

%
%

\newpage

\newpage

\end{document}